%% file: pub-nocomments.tex
\documentclass[aps,plb,twocolumn,showpacs,superscriptaddress,groupedaddress]{revtex4}

\usepackage{graphicx}%
\usepackage{dcolumn}%
\usepackage{bm}%
\usepackage{amssymb,amsmath}   %
\usepackage{epsfig}
\usepackage{psfrag}
\usepackage{xspace}
\usepackage{multirow}
\usepackage{lscape}
\usepackage{comment}

\newcommand{\sigmatt}{\mbox{${\sigma}_{t\bar{t}}$}\xspace}

\newcommand{\dzero}     {D0}
\newcommand{\ttbar}     {\mbox{$t\bar{t}$}\xspace}

\newcommand{\ppbar}     {\mbox{$p\bar{p}$}\xspace}

\newcommand{\pythia}    {\mbox{\textsc{pythia}}}
\newcommand{\alpgen}    {\mbox{\textsc{alpgen}}}
\newcommand{\mcatnlo}    {\mbox{\textsc{mc@nlo}}}
\newcommand{\geant}     {\mbox{\textsc{geant}}}

\newcommand{\met}{\mbox{\ensuremath{E\kern-0.6em\slash_T}}\xspace}
\newcommand{\metx}{\mbox{\ensuremath{E\kern-0.6em\slash_x}}\xspace}
\newcommand{\mety}{\mbox{\ensuremath{E\kern-0.6em\slash_y}}\xspace}
\newcommand{\sigmet}{\ensuremath{\sigma_{\mbox{\ensuremath{E\kern-0.6em\slash_T}}}}\xspace}

\newcommand{\ttfull}{\mbox{$t\bar{t} \rightarrow W^{+}bW^{-}\bar{b} \rightarrow \ell^{+} \nu b \ell^{-} \bar{\nu} \bar{b}$}}
\newcommand{\ttolep}{\mbox{$t \rightarrow Wb \rightarrow \ell \nu_{\ell} b$}}

\newcommand{\herwig}    {\mbox{\textsc{herwig}\xspace}}

\newcommand{\ttb}{\mbox{$t\bar{t}$}\xspace}

\begin{document}
\hspace{5.2in} \mbox{Fermilab-Pub-11/052-E}

\title{Measurement of spin correlation in $\boldsymbol{t\bar{t}}$ production using dilepton final states}

\input author_list.tex       %
\date{March 9, 2011}

\begin{abstract} %

We measure the correlation between the spin of the top quark and the
spin of the anti-top quark in \ttfull\ final states produced in
\ppbar\ collisions at a center of mass energy $\sqrt{s}=1.96$~TeV,
where $\ell$ is an electron or muon. The data correspond to an
integrated luminosity of $5.4\text{~fb}^{-1}$ and were collected with
the \dzero\ detector at the Fermilab Tevatron collider. The
correlation is extracted from the angles of the two leptons in the $t$
and $\bar{t}$ rest frames, yielding a correlation strength $C=
0.10^{+0.45}_{-0.45}$, in agreement with the NLO QCD prediction within
two standard deviations, but also in agreement with the no
correlation hypothesis.
\end{abstract}


\maketitle

\section{Introduction \label{sec:Introduction}}
Measurements of top quark properties play an important role in testing
the standard model (SM) and its possible extensions. While top and
anti-top quarks are unpolarized in $t\bar{t}$ production at hadron
colliders, the orientation of their spins are
correlated~\cite{Barger:1988jj}. The SM also predicts that top quarks
decay before the correlation between the direction of the spin of the
$t$ and $\bar{t}$ quark can be affected by
fragmentation~\cite{Bigi:1986jk}. This contrasts with the
longer-lived lighter quarks, for which the spins become decorrelated by
strong interactions before they 
decay~\cite{Falk:1993rf}. The orientation of the spin of the top
quark is therefore reflected in its decay products.

The charged leptons from the \ttolep\ decays are the probes with the
highest sensitivity to the direction of the $t$ quark spin (if not stated otherwise, charge
conjugated states are implied through out the paper). Therefore the
final state in which both $W$ bosons from $t$ quarks decay to leptons,
referred to as dilepton final state, is ideal for measurements of the
correlation between the spins of pair-produced top and anti-top quarks
and thus to test the SM~\cite{spin_theory,Brandenburg:2002xr}.

The observation of spin correlation as expected in the SM would
indicate that the top quark decays before the spin direction is
affected by its fragmentation and therefore provides an upper limit on
the lifetime of the top quark. This can be related to the
Cabibbo-Kobayashi-Maskawa matrix element $V_{tb}$ without assumptions
about the number of quark generations~\cite{Stelzer:1995gc}. Scenarios
beyond the
SM~\cite{Bernreuther:1997gs,Bigi:1986jk,Bigi:2nd,Jezabek:1994zv,Goldstein:1992xp,Bernreuther:2003xj}
predict different production and decay dynamics for the top quark,
which could affect the spin correlation.

There is a recent measurement of \ttb\ spin correlation in
semileptonic final states, in which one $W$ boson decays to leptons
and the other to quarks, by the CDF Collaboration~\cite{cdf_paper} in
$4.3\text{~fb}^{-1}$ of \ppbar\ collisions at $\sqrt{s}=1.96$~TeV
which agrees with the SM prediction. There is also an earlier
measurement analyzing dilepton final states by the \dzero\
Collaboration using an integrated luminosity of $125\text{~pb}^{-1}$
of
\ppbar\ collisions at $\sqrt{s}=1.8$~TeV~\cite{D0RunI}. However, the expected
sensitivity of both previous measurements was not high enough to distinguish
between a hypothesis of no correlation and the correlation predicted
in the SM. In both measurements a different sign convention than the
one in this article was used~\cite{signconvention}.

In this letter, we measure the strength of the $t\bar{t}$ spin
correlation $C$ from a differential angular distribution involving the
angles between the flight direction of the two decay leptons in the
rest frames of their respective $t$ quarks and the spin quantization
axis (defined below).  Top quarks are assumed to decay as predicted by
the SM.  We analyze the dilepton channels which correspond to decays
of the $W$ bosons (from $t$ and $\bar{t}$ quark decays) into an
electron and electron neutrino, a muon and a muon neutrino or a tau
lepton and a tau neutrino if the tau decays leptonically. The analysis
is performed using $5.4\text{~fb}^{-1}$ of \ppbar\ collisions at
$\sqrt{s}=1.96$~TeV collected with the \dzero\ detector at the
Fermilab Tevatron collider.

To reduce the dependence on the signal normalization, we extract the
spin correlation simultaneously with the \ttbar\ production cross
section (\sigmatt). In the SM, \sigmatt is predicted to a
precision of
($6$--$8$)\%~\cite{SMtheory_A,SMtheory_M,SMtheory_K,SMtheory_K2,SMtheory_C}. Many
models of physics beyond the SM predict effects in the top quark
sector that can affect both the top quark production rate and the spin
correlation. For example, in supersymmetric models~\cite{SUSY}, pair
production of scalar top quarks decaying into a $b$ quark, an electron
or muon, and a scalar neutrino~\cite{D0stop} would affect  the
measured values of both \sigmatt\ and $C$ in dilepton final states.

\section{Observable \label{sec:observables}}

\begin{figure}[h]
\begin{center}
\includegraphics[width=0.5\textwidth]{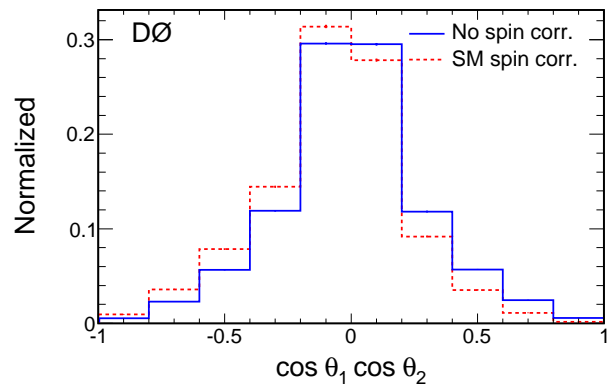}
\end{center}
\begin{center}
    \caption{\label{obs_parton} The distribution in
    $\cos\theta_{1}\cos\theta_{2}$ for a \ttb\ sample including the NLO
    QCD spin correlation ($C=0.777$) (dashed line) and with no spin
    correlation ($C=0$) (solid line) at the parton level, generated
    using \mcatnlo~\cite{mcatnlo}.}
\end{center}
\end{figure}

The \ttb\ spin correlation strength $C$ is obtained from the
distribution~\cite{spin_theory}
\begin{equation}
\label{eq:coscos}
\frac{1}{\sigma} \frac{d^{2}\sigma}{d\cos\theta_1 d\cos\theta_2} =
\frac{1}{4} ( 1 - C \cos\theta_1 \cos\theta_2 ) \,\, ,
\end{equation}
where $\sigma$ denotes the cross section, and $\theta_1$ and
$\theta_2$ are the angles between the direction of flight of the decay
leptons $\ell^+$ and $\ell^-$ in the $t$ and $\bar{t}$ rest frames and
the spin quantization axis~\cite{wignerrotation}.  These angles are
chosen because the sensitivity to the spin correlation is largest when
the decay products are down-type
fermions~\cite{Jezabek:1994zv,spin_theory,Brandenburg:2002xr}.  $C$ is
a parameter between $-1$ and $1$ that depends on the quantization axis
used and determines the magnitude of the \ttb\ spin correlation. For
the Tevatron, it has been shown in~\cite{spin_theory} that an almost
optimal choice of quantization axis is given by the direction of the
beam.  At tree level in quantum chromodynamics (QCD), $C$ represents
the number of events where the $t$ and $\bar{t}$ spins are parallel
minus the number of events where they are anti-parallel, normalized by
the total number of events. The case of all spins being
(anti-)parallel corresponds to $C=1$ ($C=-1$), whereas an equal
mixture of parallel and anti-parallel would give $C=0$.  Choosing the
beam momentum vector as the quantization axis,
$C=0.777^{+0.027}_{-0.042}$ is predicted at NLO in
QCD~\cite{spin_theory}. This calculation uses the CTEQ6.1M parton
distribution functions (PDF) with both the factorization and
renormalization scale set to the top quark mass ($m_t$). The
uncertainty reflects the variation of this scale from $m_t/2$ to
$2m_t$.  Figure~\ref{obs_parton} shows the $\cos\theta_1\cos\theta_2$
distribution calculated with spin correlation ($C=0.777$) and without
spin correlation ($C=0$). A non-vanishing spin correlation leads to an
asymmetry in the distribution.

\section{D0 Detector}
The D0 detector~\cite{d0det} contains a tracking system, a
calorimeter, and a muon spectrometer. The tracking system consists of
a silicon microstrip tracker (SMT) and a central fiber tracker, both
located inside a 1.9~T superconducting solenoid.  The design provides
efficient charged-particle tracking in the pseudorapidity region
$|\eta_{\text{det}}| < 3$~\cite{eta}. The SMT provides the capability
of reconstructing the \ppbar\ interaction vertex (PV) with a precision of
about 40 $\mu$m in the plane transverse to the beam direction and a
determination of the impact parameter of any track relative to the
PV~\cite{ip} with a precision between 20 and 50 $\mu \text{m}$,
depending on the number of hits in the SMT.  The calorimeter has a
central section (CC) covering $|\eta_{\text{det}}|<1.1$ and two end
calorimeters (EC) extending coverage to $|\eta_{\text{det}}|\approx
4.2$.  The muon system surrounds the calorimeter and consists of three
layers of tracking detectors and of scintillators covering
$|\eta_{\text{det}}|<2$~\cite{muon_detector}.  A 1.8~T toroidal iron
magnet is located outside the innermost layer of the muon detector.
The luminosity is calculated from the rate of {\mbox{$p\bar p$}}\
inelastic collisions, measured with plastic scintillator arrays which
are located in front of the EC cryostats.

The D0 trigger is based on a three-level pipeline system. The first
level consists of hardware and firmware components, the second level
combines information from different detectors to construct simple physical
quantities, while the software-based third level processes the full
event information using  simplified reconstruction algorithms
\cite{matrixmethod}.

\section{Event Selection\label{sec:Selection}}
The selections for the $\ell\ell$ ($ee$, $e\mu$, and $\mu\mu$) decay
channels follow those described in Ref.~\cite{Abazov:2009si}.
Electrons are defined as clusters of calorimeter cells for which: (i)
the energy deposited in the electromagnetic section of the calorimeter
is $>90$\% of the total cluster energy, (ii) the energy is
concentrated in a narrow cone and is isolated from other energy
depositions, (iii) the spatial distribution of the shower is
compatible with that of an electron and matches a track emanating from
the PV. To further reduce background, we use the tracking system and
calorimeter information to form a likelihood discriminant that
enhances the efficiency to select real and to reject fake
electrons. Electrons that fulfill criteria (i) to (iii) are referred
to as ``loose'' electrons, while those that also fulfill the
likelihood criterion are referred to as ``tight'' electrons. Both
central~($|\eta_{\text{det}}|<1.1$) and
forward~($1.5<|\eta_{\text{det}}|<2.5$) electrons are considered in
the analysis.

Muons are defined using tracks reconstructed in the three layers of
the muon system, with a matching track found in the central tracking
system. To reduce background, the sum of the track $p_{T}$ in
a cone of size ${\cal R}=\sqrt{(\Delta y)^2 + (\Delta \phi)^2}=0.5$
around the muon axis must be $<15$\% of the muon $p_T$. We also
require the sum of calorimeter cell energies in an annulus of radius
$0.1<\Delta{\cal R}<0.4$ relative to the axis of the muon track to be
$<15$\% of the muon $p_T$. Muon candidates originating from top quark
decays are required to have a distance of closest approach of the muon
track with respect to the PV $<0.2$~cm for a muon track
without a hit in the SMT, and $<0.02$~cm for a muon track with a
hit in the SMT. Furthermore, muons must satisfy
$|\eta_{\text{det}}|<2$.

Jets are reconstructed with a mid-point cone algorithm~\cite{jetcone}
with radius ${\cal R}=0.5$. Jet energies are corrected for calorimeter
response, additional energy from noise, pileup, and multiple \ppbar\
interactions in the same bunch crossing, and out-of-cone shower
development in the calorimeter.

Jets are required to contain three or more tracks originating from the
PV within each jet cone. The high instantaneous luminosity achieved by
the Tevatron leads to a significant background contribution from
additional \ppbar\ collisions within the same bunch crossing. The
track requirement removes jets from such collisions and is only
necessary for data taken after the initial 1~fb$^{-1}$ data set. The
missing transverse energy (\met) is defined by the magnitude and
direction opposite to the vector sum of all significant transverse
energies deposited in calorimeter cells. The transverse energy of
muons and corrections made to electron and jet energies are taken into
account. A more detailed description of object reconstruction can be
found in~\cite{Abachi:2007:prd}.

To select top quark pair events, we require two isolated, oppositely
charged leptons with $p_T>15$~GeV for all channels.  At least two jets
with $p_T>20$~GeV and $|\eta_{\text{det}}|<2.5$ are required.  The
final selection in the $e\mu$ channel requires that $H_T$ (defined as
the scalar sum of the leading lepton $p_T$ and the $p_T$ of each of
the two most energetic jets) be greater than 110~GeV.  This
requirement allows rejection of the largest backgrounds for this final
state arising from $Z/\gamma^{\ast} \rightarrow \tau^+ \tau^-$ and
diboson production.  To further reject $Z/\gamma^{\ast}$ background,
where \met\ arises from mismeasurement, we compute for each $ee$ and
$\mu\mu$ event a \met\ significance likelihood based on the
\met\ probability distribution calculated from the \met\ and the
lepton and jet energy resolutions. We require this quantity to exceed
5. We find that only in the $\mu\mu$ channel an additional cut on
missing transverse energy is beneficial to increase the signal purity,
and therefore require $\met > 40$~GeV for $\mu\mu$ final states.

\section{Signal and background modeling\label{sec:modeling}}
Signal and background processes are modeled with a combination of
Monte~Carlo (MC) simulation and data.  Top quark pair production is
simulated using the \mcatnlo~\cite{mcatnlo} generator assuming
$m_{t}=172.5$~GeV. Events are processed through
\herwig~\cite{herwig} to simulate fragmentation, hadronization and
decays of short-lived particles. 
We generate \ttb\ MC samples both with and without the expected spin
correlation, as both options are available in
\mcatnlo.
To cross check, we also use the
\alpgen~\cite{alpgen} matrix-element generator interfaced to
\pythia~\cite{pythia} to simulate parton showering.
In all three cases events are processed through a full detector
simulation using \geant~\cite{geant}.  The MC events are overlaid with
data events from random bunch crossings to model the effects of
detector noise and additional \ppbar\ interactions.  The same
reconstruction programs are then applied to data and MC events.
Lepton trigger and identification efficiencies as well as lepton
momentum resolutions are derived from $Z/\gamma^{\ast}\rightarrow
\ell^+\ell^-$ data. Jet reconstruction efficiencies and the jet
resolutions are adjusted to the values measured in data, and \met\ is
recalculated accordingly.

Sources of background arise from the production of electroweak bosons
that decay into charged leptons. In the $ee$, $e\mu$, and $\mu\mu$
channels, the dominant backgrounds are Drell-Yan processes: (i)
$Z/\gamma^{\ast}\rightarrow e^+e^-$, (ii) $Z/\gamma^{\ast}
\rightarrow \tau^+\tau^- \rightarrow \bar{\nu}
\ell^+ \nu \nu \ell^- \bar{\nu}$, and (iii) $Z/\gamma^{\ast} \rightarrow
\mu^+\mu^-$, and from diboson production  
($WW, WZ$ and $ZZ$) when the boson decays lead to two charged leptons
in the final state.  Other backgrounds can be attributed to jets
mimicking electrons, muons from semileptonic decays of $b$ quarks,  in-flight decays of pions or kaons in a jet, and misreconstructed
\met.

The selection efficiencies for the $Z/\gamma^{\ast}$ background are
estimated from MC samples generated by \alpgen\ interfaced with
\pythia, while for diboson production they are estimated using
\pythia. The diboson processes are normalized to the next-to-leading
order (NLO) inclusive cross section~\cite{mcfm}. The $Z/\gamma^{\ast}$
processes are normalized to the next-to-next-to-leading order (NNLO)
inclusive cross section~\cite{hamberg} in the $e\mu$
channel. In the other two channels we normalize the MC expectation to
the dilepton invariant mass distribution near the $Z$ boson peak.  The
$Z$ boson $p_T$ distribution predicted by \alpgen\ is observed to
agree poorly with the data, therefore a reweighting has been performed
for samples of different jet multiplicities derived from $Z
\rightarrow e^+e^-$ data events.

Before making requirements on $\met$ or its significance, the
$Z/\gamma^{\ast}$ background dominates in the $ee$ and $\mu\mu$
channels. Although these events do not contain high-$p_T$ neutrinos,
the $Z/\gamma^{\ast} \rightarrow \ell^+
\ell^-$ events can have large $\met$ from
mismeasurement or poor resolution in $\met$.

Two instrumental backgrounds are modeled using data. In the $ee$ and
$e\mu$ channels, background from fake electrons arises from jets
comprising an energetic $\pi^0$ or $\eta$ and an overlapping track. The
contribution from this source of background is estimated by fitting to
the observed distribution of an electron-likelihood discriminant in
the data, as described in~\cite{Abachi:2007:prd}. The dependence of
the electron likelihood is determined for true electrons from a pure
$Z/\gamma^{\ast}$ data sample, while the electron likelihood for
background is determined using a sample dominated by false
electrons. In the $e\mu$ and $\mu\mu$ channels, muons produced in jets
that fail to be reconstructed provide muons that appear isolated. We
measure the fraction ($f_{\mu}$) of muons that appear isolated using a
dimuon control sample dominated by false isolated muons. The
contribution from events with misidentified isolated muons is given by
the number of events in a like-sign dilepton sample (without
imposing an isolation requirement) multiplied by the measured
$f_{\mu}$ defined above.

\section{Reconstruction of Event Kinematics\label{sec:reco}}

The calculation of the angular correlation described in
Sec.~\ref{sec:observables} requires boosting the 4-momenta of the
charged leptons 
 back into the $t$ or $\bar{t}$ quark rest frames. Every 
event must therefore be fully reconstructed. This is performed using
the neutrino weighting method, devised originally for measuring $m_t$
in the dilepton channel~\cite{nuWtmass}. The only difference in our
procedure is that instead of calculating a weight distribution as a
function of the hypothesized $m_t$, we weight the distribution as a
function of $\cos\theta_1\cos\theta_2$.

In the dilepton final state, the momenta of the two charged leptons,
two neutrinos, and two $b$-quark jets are specified by eighteen
components of momentum. We measure twelve of these from the observed
leptons and jets. Four additional constraints are provided by
requiring the two lepton-neutrino combinations to yield $M_W$, and the
two $W$-boson+$b$-jet combinations to yield $m_t$ (which we assume to
be $172.5$~GeV for $t$ and $\bar{t}$). Two additional quantities must
be specified to fully reconstruct the event kinematics.

To obtain the two missing quantities, we sample the pseudorapidity
distributions $\eta_1$ and $\eta_2$ for the two neutrinos into ten
bins each using \ttb\ MC. These distributions are found to be
independent of \ttb\ spin correlation. The bin ranges are chosen in
steps of equal probability. We use the median of each bin to resolve
the kinematics of the event, with up to two solutions for each of the
neutrino transverse momenta.

The measured value of the \met\ is then used to assign a weight $w$ to
each of the solutions for each assumed set of $\eta$ values;
specifically, for a given $\eta_1$ and $\eta_2$, we calculate the
\met\ in the reconstructed event and compare it to the measured \met\
as follows:
\begin{eqnarray}
w=\exp\left[-\, \frac{\left(\metx^{calc}-\metx \right)^{2}+\left(\mety^{calc}-\mety \right)^{2}}{2\sigmet^2}
\right]  \,\, ,
\end{eqnarray}
where \sigmet is the resolution of the $x$ component of the \met
(taken to be the same as that on the $y$
component)~\cite{nuWtmass}. This assigns a higher weight ($w$) to
$\eta_{1,2}$ pairs that are consistent with the observed \met.

Since it is not possible to unambiguously associate a jet to the
correct top quark, all combinations are tried. This increases the
possible number of solutions per event from four to eight.

Detector resolutions are accommodated in the weight calculation as
follows. For each configuration of a MC event, we simulate the effect
of the detector resolution by repeating the calculation 150 times with
the measured jet and lepton momenta smeared independently according to
the detector response. The resolutions in lepton energies are assumed
to be Gaussian and the resolution of the jets is modeled using the sum
of two Gaussian distributions~\cite{alex_thesis}. The 150
resolution-smeared weights are averaged, therefore smoothing the
weight distribution of an event. This number provides stable and
smooth weight distributions, with acceptable computation times. For
data, the number of smearings is increased to 1000 to ensure that the
result does not depend on statistical fluctuations of the smearing.
\begin{table}[h]
\begin{center}
\caption{\label{tab:nwt-efficiencies}
Probability that the neutrino weighting procedure yields a valid
solution for different classes of events. The $Z\rightarrow\tau\tau$,
$Z\rightarrow\mu\mu$ and $Z\rightarrow ee$ backgrounds are shown
combined, as are the diboson and instrumental backgrounds. In the last
column we give the value observed in data.  The statistical
uncertainties are $\approx 1\%$.}
\begin{tabular}{cccccc}
\hline
\hline
 $t\bar{t}$ & $Z$ & Diboson & Instrumental & Total & Observed\\
 $0.96$ & $0.82$ & $0.90$   & $0.82$       & $0.92$ & $0.91$\\
\hline
\hline
\end{tabular}
\end{center}
\end{table}
Events without any solution are ignored in the analysis.  The
probabilities that the reconstruction of the full event kinematics
provides a valid solution for \ttb\ and background events are given in
Table~\ref{tab:nwt-efficiencies}. Table~\ref{tab:yields} summarizes
the predicted background and the observed number of events in data,
together with the number of expected \ttb\ events using the \ttb\
cross section measured in this analysis. For each event with a
solution the $w$ distribution is normalized to unity, and the mean of
the $w$ distribution is used as 
estimator for the true value of $\cos\theta_1\cos\theta_2$. The
correlation coefficient between our estimator and the true value of
$\cos\theta_1\cos\theta_2$ is about 0.5.
\begin{table*}[ht]
\begin{center}
\begin{minipage}{5.5 in}
\caption{\label{tab:yields}Yields for events with a solution in the
neutrino weighting procedure. The number of \ttbar\ events is calculated using
$\sigmatt =7.92$~pb as measured in this analysis. The
$Z\rightarrow\tau\tau$, $Z\rightarrow\mu\mu$, and $Z\rightarrow ee$
backgrounds are shown combined, as are the different instrumental
backgrounds. Uncertainties include statistical and systematic
contributions. }
\begin{tabular}{ccccccc}
\hline
\hline
& $t\bar{t}$ & $Z$ & Diboson & Instrumental & Expected & Observed\\
Number of events & $324^{+28}_{-28}$ & $75^{+13}_{-13}$ & $17^{+3}_{-3}$  & $23^{+4}_{-4}$ & $439^{+36}_{-36}$ & 441\\
\hline
\hline
\end{tabular}
\end{minipage}
\end{center}
\end{table*}

\section{Templates\label{sec:templates}}

\begin{figure}[h]
\label{fig:mean-template}
\begin{centering}
\includegraphics[width=0.5\textwidth]{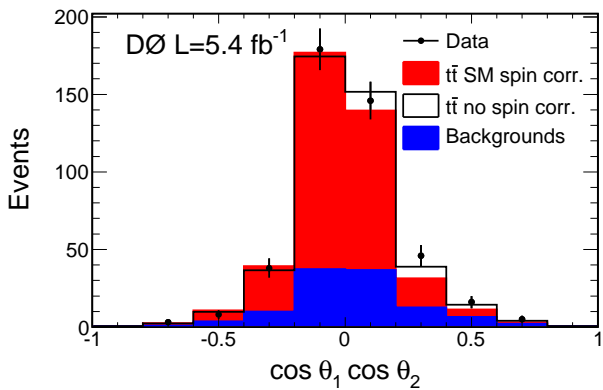}
\par\end{centering}

\caption{\label{fig:Background-and-signal-templates_channels}
(Color online) The distribution in $\cos\theta_{1}\cos\theta_{2}$ for
the entire dilepton event sample. The summed \ttb\ signal, including
NLO QCD spin correlation ($C=0.777$) (red) and all backgrounds (blue)
are compared to data. The open histogram is the \ttb\ prediction
without spin correlation ($C=0$). The slight asymmetry in the
$\cos\theta_{1}\cos\theta_{2}$ distribution does not bias the
measurement.}

\end{figure}

Templates of the $\cos\theta_1\cos\theta_2$ distributions are
generated using MC events for different values of $C$ and then
compared to
data. Figure~\ref{fig:Background-and-signal-templates_channels} shows
the distribution for $\cos\theta_1\cos\theta_2$ for background,
\ttb\ signal with NLO QCD spin correlation, and the prediction
for \ttb\ signal without spin correlation.
Different fractions of events without and with SM spin correlation can
be used to generate \ttb\ samples of different true values of
$C$. Templates are formed from the sum of \ttb\ signals of different
$C$ values and contributions from backgrounds, as a function of
$\cos\theta_1\cos\theta_2$. As the ratio of signal to background is
different in $ee$, $e\mu$ and $\mu\mu$ final states,
we analyze each channel separately. In each
channel, we use eight bins of equal size over the range
$\left[-0.4,0.4\right]$, and additionally one bin each for the range
$\left[-1,-0.4\right]$ and $\left[0.4,1\right]$. The 
$\cos\theta_{1}\cos\theta_{2}$ distributions in data are compared with
these templates to extract the best measured value for $C$
($C_{\mathrm{meas}}$).

\section{Fit to Templates and Systematic Uncertainties \label{sec:TemplateFit}}
We perform a binned maximum likelihood fit to extract the measured
value $C_{\mathrm{meas}}$.  We maximize the likelihood function
\begin{equation}
{\cal L} = \prod_{i} {\cal P}(n_{i}, m_{i})
\times \prod_{k=1}^{K} {\cal G}(\nu_k;0,{\text{SD}}_k) \,,
\label{eq:mlikeli}
\end{equation}
where ${\cal P}(n,m)$ represents the Poisson probability
to observe $n$ events when $m$ events are expected. The first product
runs over all the bins $i$ of the templates of all
channels. Systematic uncertainties are taken into account by
parameters $\nu_k$, where each independent source of
systematic uncertainty $k$ is modeled as a Gaussian probability
density function, ${\cal G}\left(\nu;0,{\text{SD}}\right)$, with zero
mean and width corresponding to one standard deviation (SD) in the
uncertainty of that parameter. Correlations among systematic
uncertainties between channels are taken into account by using the
same parameter for the same sources of uncertainty.  The predicted
number of events in each bin is the sum of the predicted number of
background and expected \ttb\ events and depends on $C$.

We consider both systematic uncertainties which affect only
normalization factors and those which alter the differential
distribution of $\cos\theta_{1}\cos\theta_{2}$. Uncertainties derived
from differences between data and MC for the jet energy scale, jet
energy resolution, jet identification, and from theoretical
uncertainties on PDFs, background modeling, and the choice of $m_t$
are taken as differential in
$\cos\theta_{1}\cos\theta_{2}$. Systematic uncertainties affecting the
overall signal efficiency and the normalization of backgrounds include
lepton identification, trigger requirements, uncertainties on the
normalization of background, the uncertainty on the luminosity, MC
modeling, and the determination of instrumental background. We also
include an uncertainty on the templates because of limited statistics
in the MC samples. We estimate the latter from 1000
pseudo-experiments, where we randomly vary each bin in the templates
within the statistical uncertainty of the MC and repeat the
measurement on data.

The statistical and systematic uncertainties on $C_{\mathrm{meas}}$
are listed in Table~\ref{tab:systematics_C}.  We evaluate the size of
the individual sources of systematic uncertainty by setting all
parameters for the systematic uncertainties to their fitted mean value
and calculate the impact of the upward and downward one standard
deviation uncertainty on the fitted parameter on the measured
quantities $C_{\mathrm{meas}}$ and $\sigma_{t\bar{t}}$.

\begin{table*}[ht]
\begin{center}
\begin{minipage}{5.5 in}
\caption{\label{tab:systematics_C}Summary of uncertainties on $C_{\mathrm{meas}}$. 
} 
\vspace{0.2cm}
\setlength{\tabcolsep}{9pt}
{
\renewcommand{\arraystretch}{1.1} 
\begin{tabular}{ccc} \hline \hline
 Source  &        $+$SD    &   $-$SD   \\ \hline
                   Muon identification & 0.01 & $-0.01$ \\ 
              Electron identification and smearing & 0.01 & $-0.01$ \\ 
                                     PDF & 0.02 & $-0.01$ \\ 
                                   Top Mass & 0.01 & $-0.01$ \\ 
                                          Triggers & 0.02 & $-0.02$ \\ 
                       Opposite charge requirement & 0.00 & $-0.00$ \\ 
                                  Jet energy scale & 0.01 & $-0.01$ \\ 
                Jet reconstruction and identification & 0.06 & $-0.06$ \\ 
                                    Normalization & 0.02 & $-0.02$ \\ 
                            Monte Carlo statistics & 0.02 & $-0.02$ \\ 
                Instrumental background & 0.00 & $-0.00$ \\ 
                              Background Model for Spin & 0.03 & $-0.04$ \\ 
                                    Luminosity & 0.03 & $-0.03$ \\ 
                                             Other & 0.01 & $-0.01$ \\ 
                 Template statistics for template fits & 0.07 & $-0.07$ \\ 
                                 \hline  
                   Total systematic uncertainty & 0.11 & $-0.11$ \\ \hline
   Statistical uncertainty    & 0.38 & $-0.40$  \\ \hline \hline
 \end{tabular}
}
\end{minipage}
 \end{center}
\end{table*}
\section{Result\label{sec:result}}

\begin{figure}[h]
\begin{centering}
\includegraphics[width=0.5\textwidth]{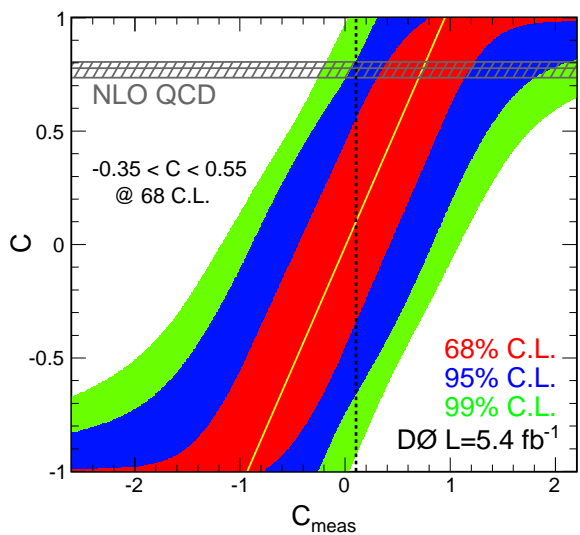}
\par\end{centering}

\caption{\label{fig:fc_c}
(Color online) The 68\% (inner), 95\% (middle), and 99\% (outer)
C.L. bands of $C$ as a function of $C_{\mathrm{meas}}$ from likelihood
fits to MC events for all channels combined.  The yellow line
indicates the most probable value of $C$ as a function of
$C_{\mathrm{meas}}$, and represents the calibration of the method.
The vertical dashed black line depicts the measured value
$C_{\mathrm{meas}}=0.10$. The horizontal band indicates the NLO QCD
prediction of $C=0.777^{+0.027}_{-0.042}$.}
\end{figure}

To estimate the expected uncertainty of the result, ensembles of MC
experiments are generated for different values of $C$, and the maximum
likelihood fit is repeated, yielding a distribution of
$C_{\mathrm{meas}}$ for each $C$.  Systematic uncertainties are
included in this procedure, taking into account correlations between
channels. We then apply the ``ordering principle'' for ratios of
likelihoods~\cite{fc_limit} to the MC distributions of 
$C_{\mathrm{meas}}$ and generated $C$.
Due to fluctuations in the data the best fit value for
$C_{\mathrm{meas}}$ can lie outside the physical region,
however, a physically meaningful value for $C$ can be extracted for any
value of $C_{\mathrm{meas}}$ by using Fig.~\ref{fig:fc_c}. For the
SM expectation of $C=0.777$ we expect to exclude values below $-0.06$
at the 95\% C.L. From the maximum likelihood fit to data we obtain
$C_{\mathrm{meas}} = 0.10^{+0.42}_{-0.44} \, ({\text{stat+syst}}) $,
which is shown in Fig.~\ref{fig:fc_c}. We transform
$C_{\mathrm {meas}}$ into 
\begin{equation}
C=0.10^{+0.45}_{-0.45} \, ({\text{stat+syst}}) \,\,,
\end{equation}
and extract a $95$\% C.L. region of probability for $C$ as
$\left[-0.66,0.81\right]$. Our result
is within two standard deviations of the NLO QCD prediction of
$C=0.777^{+0.027}_{-0.042}$ but also compatible with the
no-correlation hypothesis.

The simultaneously extracted \ttbar\ cross section is found to be
\begin{equation}
\sigmatt = 7.92 ^{+1.07}_{-0.93} \, ({\text{stat + syst}}) \ \, {\text{pb}}
\end{equation}
for $m_t=172.5$~GeV, which agrees with the SM prediction
of $\sigmatt = 7.46 ^{+0.48}_{-0.67} \ \, {\mathrm{pb}}$~\cite{SMtheory_M}.

\section{Conclusion}

We have measured the spin correlation between top and anti-top quarks
in $t\bar{t}$ production using a differential angular distribution in
lepton decay angles. The correlation coefficient characterizing the
degree of spin correlation is found to be $C =
0.10^{+0.45}_{-0.45}$. This is the most precise result obtained from
the analysis of top quark pair production in dilepton final states.
Since the measured value of $C$ agrees with the SM prediction of
$C=0.777^{+0.027}_{-0.042}$ in NLO QCD within two standard deviations,
there is no significant hint for anomalous production or decay of top
quark pairs.

\section*{Acknowledgments}
We wish to thank W.~Bernreuther, S.~J.~Parke, and P.~Uwer
for fruitful discussions regarding this analysis.
We thank the staffs at Fermilab and collaborating institutions, 
and acknowledge support from the 
DOE and NSF (USA);
CEA and CNRS/IN2P3 (France);
FASI, Rosatom and RFBR (Russia);
CNPq, FAPERJ, FAPESP and FUNDUNESP (Brazil);
DAE and DST (India);
Colciencias (Colombia);
CONACyT (Mexico);
KRF and KOSEF (Korea);
CONICET and UBACyT (Argentina);
FOM (The Netherlands);
STFC and the Royal Society (United Kingdom);
MSMT and GACR (Czech Republic);
CRC Program and NSERC (Canada);
BMBF and DFG (Germany);
SFI (Ireland);
The Swedish Research Council (Sweden);
and
CAS and CNSF (China).

\newpage

\end{document}

%% file: author_list.tex
\affiliation{Universidad de Buenos Aires, Buenos Aires, Argentina}
\affiliation{LAFEX, Centro Brasileiro de Pesquisas F{\'\i}sicas, Rio de Janeiro, Brazil}
\affiliation{Universidade do Estado do Rio de Janeiro, Rio de Janeiro, Brazil}
\affiliation{Universidade Federal do ABC, Santo Andr\'e, Brazil}
\affiliation{Instituto de F\'{\i}sica Te\'orica, Universidade Estadual Paulista, S\~ao Paulo, Brazil}
\affiliation{Simon Fraser University, Vancouver, British Columbia, and York University, Toronto, Ontario, Canada}
\affiliation{University of Science and Technology of China, Hefei, People's Republic of China}
\affiliation{Universidad de los Andes, Bogot\'{a}, Colombia}
\affiliation{Charles University, Faculty of Mathematics and Physics, Center for Particle Physics, Prague, Czech Republic}
\affiliation{Czech Technical University in Prague, Prague, Czech Republic}
\affiliation{Center for Particle Physics, Institute of Physics, Academy of Sciences of the Czech Republic, Prague, Czech Republic}
\affiliation{Universidad San Francisco de Quito, Quito, Ecuador}
\affiliation{LPC, Universit\'e Blaise Pascal, CNRS/IN2P3, Clermont, France}
\affiliation{LPSC, Universit\'e Joseph Fourier Grenoble 1, CNRS/IN2P3, Institut National Polytechnique de Grenoble, Grenoble, France}
\affiliation{CPPM, Aix-Marseille Universit\'e, CNRS/IN2P3, Marseille, France}
\affiliation{LAL, Universit\'e Paris-Sud, CNRS/IN2P3, Orsay, France}
\affiliation{LPNHE, Universit\'es Paris VI and VII, CNRS/IN2P3, Paris, France}
\affiliation{CEA, Irfu, SPP, Saclay, France}
\affiliation{IPHC, Universit\'e de Strasbourg, CNRS/IN2P3, Strasbourg, France}
\affiliation{IPNL, Universit\'e Lyon 1, CNRS/IN2P3, Villeurbanne, France and Universit\'e de Lyon, Lyon, France}
\affiliation{III. Physikalisches Institut A, RWTH Aachen University, Aachen, Germany}
\affiliation{Physikalisches Institut, Universit{\"a}t Freiburg, Freiburg, Germany}
\affiliation{II. Physikalisches Institut, Georg-August-Universit{\"a}t G\"ottingen, G\"ottingen, Germany}
\affiliation{Institut f{\"u}r Physik, Universit{\"a}t Mainz, Mainz, Germany}
\affiliation{Ludwig-Maximilians-Universit{\"a}t M{\"u}nchen, M{\"u}nchen, Germany}
\affiliation{Fachbereich Physik, Bergische Universit{\"a}t Wuppertal, Wuppertal, Germany}
\affiliation{Panjab University, Chandigarh, India}
\affiliation{Delhi University, Delhi, India}
\affiliation{Tata Institute of Fundamental Research, Mumbai, India}
\affiliation{University College Dublin, Dublin, Ireland}
\affiliation{Korea Detector Laboratory, Korea University, Seoul, Korea}
\affiliation{CINVESTAV, Mexico City, Mexico}
\affiliation{FOM-Institute NIKHEF and University of Amsterdam/NIKHEF, Amsterdam, The Netherlands}
\affiliation{Radboud University Nijmegen/NIKHEF, Nijmegen, The Netherlands}
\affiliation{Joint Institute for Nuclear Research, Dubna, Russia}
\affiliation{Institute for Theoretical and Experimental Physics, Moscow, Russia}
\affiliation{Moscow State University, Moscow, Russia}
\affiliation{Institute for High Energy Physics, Protvino, Russia}
\affiliation{Petersburg Nuclear Physics Institute, St. Petersburg, Russia}
\affiliation{Instituci\'{o} Catalana de Recerca i Estudis Avan\c{c}ats (ICREA) and Institut de F\'{i}sica d'Altes Energies (IFAE), Barcelona, Spain}
\affiliation{Stockholm University, Stockholm and Uppsala University, Uppsala, Sweden }
\affiliation{Lancaster University, Lancaster LA1 4YB, United Kingdom}
\affiliation{Imperial College London, London SW7 2AZ, United Kingdom}
\affiliation{The University of Manchester, Manchester M13 9PL, United Kingdom}
\affiliation{University of Arizona, Tucson, Arizona 85721, USA}
\affiliation{University of California Riverside, Riverside, California 92521, USA}
\affiliation{Florida State University, Tallahassee, Florida 32306, USA}
\affiliation{Fermi National Accelerator Laboratory, Batavia, Illinois 60510, USA}
\affiliation{University of Illinois at Chicago, Chicago, Illinois 60607, USA}
\affiliation{Northern Illinois University, DeKalb, Illinois 60115, USA}
\affiliation{Northwestern University, Evanston, Illinois 60208, USA}
\affiliation{Indiana University, Bloomington, Indiana 47405, USA}
\affiliation{Purdue University Calumet, Hammond, Indiana 46323, USA}
\affiliation{University of Notre Dame, Notre Dame, Indiana 46556, USA}
\affiliation{Iowa State University, Ames, Iowa 50011, USA}
\affiliation{University of Kansas, Lawrence, Kansas 66045, USA}
\affiliation{Kansas State University, Manhattan, Kansas 66506, USA}
\affiliation{Louisiana Tech University, Ruston, Louisiana 71272, USA}
\affiliation{Boston University, Boston, Massachusetts 02215, USA}
\affiliation{Northeastern University, Boston, Massachusetts 02115, USA}
\affiliation{University of Michigan, Ann Arbor, Michigan 48109, USA}
\affiliation{Michigan State University, East Lansing, Michigan 48824, USA}
\affiliation{University of Mississippi, University, Mississippi 38677, USA}
\affiliation{University of Nebraska, Lincoln, Nebraska 68588, USA}
\affiliation{Rutgers University, Piscataway, New Jersey 08855, USA}
\affiliation{Princeton University, Princeton, New Jersey 08544, USA}
\affiliation{State University of New York, Buffalo, New York 14260, USA}
\affiliation{Columbia University, New York, New York 10027, USA}
\affiliation{University of Rochester, Rochester, New York 14627, USA}
\affiliation{State University of New York, Stony Brook, New York 11794, USA}
\affiliation{Brookhaven National Laboratory, Upton, New York 11973, USA}
\affiliation{Langston University, Langston, Oklahoma 73050, USA}
\affiliation{University of Oklahoma, Norman, Oklahoma 73019, USA}
\affiliation{Oklahoma State University, Stillwater, Oklahoma 74078, USA}
\affiliation{Brown University, Providence, Rhode Island 02912, USA}
\affiliation{University of Texas, Arlington, Texas 76019, USA}
\affiliation{Southern Methodist University, Dallas, Texas 75275, USA}
\affiliation{Rice University, Houston, Texas 77005, USA}
\affiliation{University of Virginia, Charlottesville, Virginia 22901, USA}
\affiliation{University of Washington, Seattle, Washington 98195, USA}
\author{V.M.~Abazov} \affiliation{Joint Institute for Nuclear Research, Dubna, Russia}
\author{B.~Abbott} \affiliation{University of Oklahoma, Norman, Oklahoma 73019, USA}
\author{B.S.~Acharya} \affiliation{Tata Institute of Fundamental Research, Mumbai, India}
\author{M.~Adams} \affiliation{University of Illinois at Chicago, Chicago, Illinois 60607, USA}
\author{T.~Adams} \affiliation{Florida State University, Tallahassee, Florida 32306, USA}
\author{G.D.~Alexeev} \affiliation{Joint Institute for Nuclear Research, Dubna, Russia}
\author{G.~Alkhazov} \affiliation{Petersburg Nuclear Physics Institute, St. Petersburg, Russia}
\author{A.~Alton$^{a}$} \affiliation{University of Michigan, Ann Arbor, Michigan 48109, USA}
\author{G.~Alverson} \affiliation{Northeastern University, Boston, Massachusetts 02115, USA}
\author{G.A.~Alves} \affiliation{LAFEX, Centro Brasileiro de Pesquisas F{\'\i}sicas, Rio de Janeiro, Brazil}
% ...
%
%
\collaboration{The D0 Collaboration\footnote{with visitors from
$^{a}$Augustana College, Sioux Falls, SD, USA,
$^{b}$The University of Liverpool, Liverpool, UK,
$^{c}$SLAC, Menlo Park, CA, USA,
$^{d}$University College London, London, UK,
$^{e}$Centro de Investigacion en Computacion - IPN, Mexico City, Mexico,
$^{f}$ECFM, Universidad Autonoma de Sinaloa, Culiac\'an, Mexico,
and 
$^{g}$Universit{\"a}t Bern, Bern, Switzerland.%
}} \noaffiliation
\vskip 0.25cm

%% file: pub-nocomments.bbl
\begin{thebibliography}{99}


%
\bibitem{Barger:1988jj}
  V.~D.~Barger, J.~Ohnemus, and R.~J.~N.~Phillips,
  %
  %
  %
  Int.\ J.\ Mod.\ Phys.\  A {\bf 4}, 617 (1989).
  %

%
\bibitem{Bigi:1986jk}
  I.~I.~Y.~Bigi, Y.~L.~Dokshitzer, V.~A.~Khoze, J.~H.~K\"uhn, and  P.~M.~Zerwas,
  %
  Phys.\ Lett.\  B {\bf 181}, 157 (1986).
  %

%
\bibitem{Falk:1993rf}
  A.~F.~Falk and M.~E.~Peskin,
  %
  Phys.\ Rev.\  D {\bf 49}, 3320 (1994).
%
  %

%
\bibitem{Brandenburg:2002xr}
  A.~Brandenburg, Z.~G.~Si, and P.~Uwer,
  %
  %
  Phys.\ Lett.\  B {\bf 539}, 235 (2002).
  %
  %

\bibitem{spin_theory}
%
  W.~Bernreuther, A.~Brandenburg, Z.~G.~Si and P.~Uwer,
  %
  %
  %
  %
  %
  Nucl.\ Phys.\  B {\bf 690}, 81 (2004).
  %
  %

%
\bibitem{Stelzer:1995gc}
  T.~Stelzer and S.~Willenbrock,
  %
  Phys.\ Lett.\  B {\bf 374}, 169 (1996).
%
  %

%
\bibitem{Bernreuther:1997gs}
  W.~Bernreuther, M.~Flesch and P.~Haberl,
  %
  %
  Phys.\ Rev.\  D {\bf 58}, 114031 (1998).
%
  %


\bibitem{Bigi:2nd}
  I.~I.~Y.~Bigi, Phys.\ Lett.\  B {\bf 175}, 233 (1986).

%
\bibitem{Jezabek:1994zv}
  M.~Jezabek and J.~H.~K\"uhn,
  %
  Phys.\ Lett.\  B {\bf 329}, 317 (1994).
%
  %

%
\bibitem{Goldstein:1992xp}
  G.~R.~Goldstein, K.~Sliwa, and R.~H.~Dalitz,
  %
  Phys.\ Rev.\  D {\bf 47}, 967 (1993).
  %
  %

%
\bibitem{Bernreuther:2003xj}
  W.~Bernreuther, M.~F\"ucker, and Y.~Umeda,
  %
  %
  Phys.\ Lett.\  B {\bf 582}, 32 (2004).
  %
  %



%
\bibitem{cdf_paper}
  T.~Aaltonen {\it et al.} [CDF Collaboration],
  %
  %
  Phys.\ Rev.\  D {\bf 83}, 031104 (2011).
  %
  %



\bibitem{D0RunI}
  B.~Abbott {\it et al.}  [D0 Collaboration],
%
%
  Phys.\ Rev.\ Lett.\  {\bf 85}, 256 (2000).
  %
  %

\bibitem{signconvention} We follow the convention in Ref.~\cite{spin_theory}.

\bibitem{SMtheory_A}
  %
  V.~Ahrens {\it et al.},
  %
  %
  J. High Energy Phys. {\bf 09}, 097 (2010);
  %
  %
  %
  V.~Ahrens {\it et al.},
  %
  Nucl.\ Phys.\ Proc.\ Suppl.\  {\bf 205-206}, 48 (2010).
  %
  %


\bibitem{SMtheory_M} S.~Moch and P.~Uwer, Phys. Rev. D {\bf 78}, 034003 (2008);
  U.~Langenfeld, S.~Moch, and P.~Uwer,
  %
  Phys.\ Rev.\  D {\bf 80}, 054009 (2009);
  %
  M.~Aliev {\it et al.},
  %
  Comput.\ Phys.\ Commun.\  {\bf 182}, 1034 (2011).
  %
  %


\bibitem{SMtheory_K} N.\ Kidonakis and R.\ Vogt, Phys.\ Rev.\ D {\bf 68}, 114014 (2003).

\bibitem{SMtheory_K2}
  N.~Kidonakis,
  %
  %
  Phys.\ Rev.\  D {\bf 82}, 114030 (2010).
  %
  %



\bibitem{SMtheory_C}
      %
        %
	M.~Cacciari {\it et al.},
        %
        %
        J. High Energy Phys. {\bf 04}, 068 (2004).
%
        %


%
\bibitem{SUSY}
  P.~Fayet and S.~Ferrara,
  %
  Phys.\ Rept.\  {\bf 32}, 249 (1977).
  %


%
\bibitem{D0stop}
  V.~M.~Abazov {\it et al.}  [D0 Collaboration],
  %
  %
  Phys.\ Lett.\  B {\bf 696}, 321 (2011).
  %
  %

\bibitem{wignerrotation} Here the $t$ and $\bar{t}$ quark rest frames
are obtained by first boosting into the \ttbar\ zero mass frame. If
the $t$ and $\bar{t}$ quark rest frames were constructed by directly
boosting from the laboratory frame, a Wigner rotation would have to be
taken into account.

%
\bibitem{mcatnlo}
  S.~Frixione and B.~R.~Webber,
%
  J. High Energy Phys. {\bf 06}, 029 (2002).
%
  %

 \bibitem{d0det}
%
  V.~M.~Abazov {\it et al.}  [D0 Collaboration],
%
  Nucl.\ Instrum.\ Methods  Phys. Res. A {\bf 565}, 463 (2006);
  M.~Abolins {\it et al.}, Nucl.\ Instrum.\ Methods Phys. Res. A {\bf 584}, 75 (2008);
  R.~Angstadt {\it et al.}, Nucl.\ Instrum.\ Methods Phys. Res. A {\bf 622}, 298 (2010).
%
  %




\bibitem{eta} The rapidity $y$ and pseudorapidity $\eta$ of a particle are defined as functions
    of the polar angle $\theta$ and velocity parameter $\beta$ as
    $y(\theta,\beta) \equiv {\frac{1}{2}}
    \ln{[(1+\beta\cos{\theta})/(1-\beta\cos{\theta})]}$ and
    $\eta(\theta) \equiv y(\theta,1)$, where $\beta$ is the ratio of a
    particle's momentum to its energy. We distinguish detector $\eta$
    ($\eta_{\text{det}}$) and collision $\eta$, where the former is defined
    with respect to the center of the detector and the latter with
    respect to the \ppbar\ interaction vertex.

\bibitem{ip} The impact parameter is defined as the distance of closest 
    approach ($d_{ca}$) of the track to the PV in the plane transverse
    to the beamline. The impact parameter significance is defined as
    $d_{ca}/\sigma_{d_{ca}}$, where $\sigma_{d_{ca}}$ is the
    uncertainty on $d_{ca}$.

 \bibitem{muon_detector}
  V.~M.~Abazov {\it et al.},
  %
  Nucl.\ Instrum.\ Meth.\  A {\bf 552}, 372 (2005).
  %
  %

\bibitem{matrixmethod}
%
  V.~M.~Abazov {\it et al.}  [D0 Collaboration],
  %
  %
  Phys.\ Rev.\  D {\bf 76}, 092007 (2007).
%
  %

%
%
\bibitem{Abazov:2009si}
  V.~M.~Abazov {\it et al.}  [D0 Collaboration],
  %
  %
  Phys.\ Lett.\  B {\bf 679}, 177 (2009).
  %
  %


\bibitem{jetcone}
  G.~C.~Blazey {\it et al.}, arXiv:hep-ex/0005012 (2000).

%
\bibitem{Abachi:2007:prd}
  V.~M.~Abazov {\it et al.}  [D0 Collaboration],
 %
  Phys.\ Rev.\ D {\bf 76}, 052006 (2007).


\bibitem{herwig}
G. Corcella {\it et al.}, J. High Energy Phys. {\bf 01}, 010 (2001). 

\bibitem{alpgen} M. L. Mangano {\it et al.}, J. High Energy Phys. {\bf 07}, 001 (2003). We use \alpgen\ version 2.11.

\bibitem{pythia} T. Sj\"ostrand {\it et al.}, Comp.
  Phys. Commun. \ {\bf 135}, 238 (2001). We use \pythia\ version 6.409.

\bibitem{geant}
  R.\ Brun and F.\ Carminati, CERN Program Library Long Writeup W5013,
  1993 (unpublished).

\bibitem{mcfm} J.M.~Campbell and R.K.~Ellis,
  Phys.\ Rev.\ D \ {\bf 60}, 113006 (1999).

\bibitem{hamberg}
 R.~Hamberg, W.~L.~van Neerven, T.~Matsuura,
  %
  Nucl.\ Phys.\  {\bf B359}, 343-405 (1991) [Erratum-ibid. B {\bf 644}, 403 (2002)].

%
\bibitem{nuWtmass}
  V.~M.~Abazov {\it et al.}  [D0 Collaboration],
  %
  Phys.\ Rev.\  D {\bf 80}, 092006 (2009);
  %
  %
  B.~Abbott {\it et al.}  [D0 Collaboration],
  %
  %
  Phys.\ Rev.\ Lett.\  {\bf 80}, 2063 (1998).
  %
  %

\bibitem{alex_thesis}
A. Grohsjean, Fermilab-Thesis-2008-92 (2008).

%
\bibitem{fc_limit}
  G.~J.~Feldman and R.~D.~Cousins,
  %
  %
  Phys.\ Rev.\  D {\bf 57}, 3873 (1998).
  %
  %

\end{thebibliography}
